\documentclass{svproc}
\usepackage{url}

\usepackage{hyperref}
\usepackage{type1cm}        
\usepackage{makeidx}         
\usepackage{graphicx}        

\usepackage{multicol}        
\usepackage[bottom]{footmisc}

\usepackage{newtxtext}       %
\usepackage[varvw]{newtxmath}       

\usepackage{bm}
\usepackage{siunitx}
\usepackage{subcaption}

\newcounter{algorithm}

\newcommand{\AEW}[1]{{\textcolor{black}{#1}}}
\newcommand{\NK}[1]{{\textcolor{black}{#1}}}

\DeclareSymbolFont{bbold}{U}{bbold}{m}{n}
\DeclareSymbolFontAlphabet{\mathbbold}{bbold}

\begin{document}

\mainmatter              
\title{Enhancing Neural Autoregressive Distribution Estimators 
for Image Reconstruction}
\titlerunning{NADE for Image Reconstruction}  
%
\author{Ambrose Emmett-Iwaniw\inst{1} \and Nathan Kirk\inst{2}}
\authorrunning{Ambrose Emmett-Iwaniw et al.} 
%
\tocauthor{Ambrose Emmett-Iwaniw and Nathan Kirk }
\institute{University of Waterloo, Waterloo ON, Canada, \\
\email{arsemmettiwaniw@uwaterloo.ca},
\and
Illinois Institute of Technology, Chicago IL, 60616, USA,\\
\email{nkirk@illinoistech.edu}}

\maketitle              

\begin{abstract}
Autoregressive models are often employed to learn distributions of image data by decomposing the $D$-dimensional density function into a product of one-dimensional conditional distributions. Each conditional depends on preceding variables (pixels, in the case of image data), making the order in which variables are processed fundamental to the model performance. In this paper, we study the problem of observing a small subset of image pixels (referred to as a pixel patch) to predict the unobserved parts of the image. As our prediction mechanism, we propose a generalized version of the convolutional neural autoregressive distribution estimation (ConvNADE) model adapted for real-valued and color images. Moreover, we investigate the quality of image reconstruction when observing both random pixel patches and low-discrepancy pixel patches inspired by quasi-Monte Carlo theory. Experiments on benchmark datasets demonstrate that, where design permits, pixels sampled or stored to preserve uniform coverage improves reconstruction fidelity and test performance.
\end{abstract}
\section{Introduction}

In statistics and machine learning, distribution estimation is a fundamental problem with wide-ranging applications. Accurately estimating the underlying distribution of data allows one to tackle a wide range of inference problems such as inpainting, the process of filling in missing parts of data. In the context of image data, this involves predicting unobserved pixels, often missing due to hardware damage or image compression. Beyond image related tasks, precise distribution estimators can also provide the foundation for more advanced modeling techniques, for example in large language models (LLMs), making them indispensable tools in the modern machine learning toolkit.

\subsection{Autoregressive Models} 

The concept of autoregression was first introduced in the early twentieth century in the context of time series analysis, which is used to model and forecast time series data. The simplest autoregressive model, the $AR(p)$ model introduced in \cite{YULE1927}, represents the value of a variable at a given time as a linear combination of its previous values, where $p$ denotes the number of lagged observations included in the model.

The transition of autoregressive models from traditional statistics to machine learning occurred as researchers recognized their potential for modeling complex, high-dimensional data. These models leverage the chain rule of probability, which allows them to factorize the joint distribution of all the variables in a distribution into a product of conditional distributions. Specifically, the probability of each variable is modeled as a function of the preceding variables. Frey \cite{frey1998graphical} developed the first autoregressive model of this kind by considering the use of logistic regression models for these conditionals. Later, \cite{bengio1999modeling} extended the conditionals to be more flexible by using a single-layer feed-forward neural network with a form of parameter sharing paving the way for more powerful and flexible approaches. This work laid the groundwork for models like the neural autoregressive distribution estimation (NADE) model developed in \cite{larochelle2011neural}, which was introduced as an efficient model primarily due to the advantage of weight sharing across all the conditional probabilities, given an arbitrary ordering of the input variables. NADE's introduction was significant and influential; it provided a way to model high-dimensional distributions using a neural network that was both computationally efficient and capable of capturing complex dependencies in the data. This approach inspired further developments, such as RNADE for real-valued data \cite{uria2013rnade} and ConvNADE for image data \cite{uria2016neural}, addressing specific limitations of earlier models. The mathematical details of NADE and ConvNADE models are given in Section \ref{sec:backgroundonNADE}.


\subsection{The Ordering of Input Variables}
In the NADE model and its variants, the ordering of input variables is important in determining the model's performance and computational efficiency. This intrinsic link stems from their autoregressive philosophy where given a target distribution function $p$ and an input vector of observations $\mathbf{x} = (x_1,\ldots,x_D)$, 
autoregressive methods use the product rule to factorize the probability density
function of a $D$-dimensional vector as a product of one-dimensional conditional distributions in a given order $o$, where $o$ is a permutation of $\{1,\dots,D\}$. That is,
\begin{equation}\label{eq:conditionals}
p(\mathbf{x}) = \prod_{d=1}^{D}p(x_{o_d} | \mathbf{x}_{o_{<d}}).
\end{equation}
The notation $x_{o_d}$ denotes the $d^{th}$ element of vector $\mathbf{x}$ under ordering $o$, $o_{<d}$ contains the first $(d-1)$ dimensions in ordering $o$ and $\mathbf{x}_{o_{<d}}$ is the corresponding $(d-1)$-dimensional sub-vector for these dimensions. Each conditional distribution is computed based on the preceding variables in the given ordering, which has significant implications for how the model captures dependencies and patterns in the data.

In \cite{uria2014deep}, the authors recognized the significance of ordering and subsequently introduced the idea of order-agnostic NADE models, i.e., a model not dependent on a specific variable ordering, $o$. One such model, known as DeepNADE, trains the model not on one specific sequence of input variables but by exposing the model to multiple random orderings of the input vector of observations during the training process. This approach encourages the model to generalize across different possible sequences of variable dependencies, leading to more robust performance. Subsequently in \cite{uria2016neural}, it was noted that in the context of image data, the DeepNADE model flattens the two-dimensional image to a one-dimensional vector, hence removing the spatial topology that contains useful information regarding which pixels likely depend on each other. To mitigate this issue and to ensure that the spatial information is not lost during the training process, the ConvNADE model was introduced as a further extension of DeepNADE that employs convolutional layers (as in convolutional neural networks) to preserve the spatial topology of the image. Importantly, while the order-agnostic training that occurs in both ConvNADE and DeepNADE enhances the flexibility and robustness during the inference stage, it also introduces additional computational complexity during the training stage
as the model must effectively learn from a much larger space of possible variable input sequences.

\subsection{Image Reconstruction}

\NK{One primary application of autoregressive modeling, and the main subject of this paper, is image reconstruction. That is, given $P$ pixels from a full $D > P$ pixel image, predict the missing $D-P$ pixels. In this work, our prediction mechanism will be a variant of the ConvNADE model adapted specifically for this image reconstruction task. Namely, instead of conditioning on one pixel at a time in the autoregressive procedure, we will instead condition on a small subset of pixels. This training and reconstruction procedure is explained more fully in Section \ref{sec:new_way}.}

\NK{There are various alternative approaches to address the image reconstruction problem, including image transformers \cite{chen2020generative,khan2022transformers} and advanced denoising autoencoders \cite{vincent2008extracting,vincent2010stacked}, such as the masked autoencoder \cite{he2022masked}, which are commonly used for pre-training models to enhance image classification tasks \cite{dong2023peco,he2022masked,Diff_images_2024}. Diffusion models have also become one of the most popular machine learning models for their superiority in image related tasks \cite{Diff_models_2021_beatGANS}. However, all of the models mentioned above demand extensive computational resources, and therefore often build upon pre-trained models requiring hundreds of hours of training with hundreds of GPUs distributed across clusters \cite{latent_diffusion_models_2022,Diff_images_2024}. As a result, they fall outside the scope of this work, which instead aims to provide a more computationally efficient solution for image reconstruction. While our method does not claim to match the reconstruction quality of these high-capacity state-of-the-art models, it offers a lightweight alternative that can feasibly be trained from scratch on a single GPU. Our proposed model also provides a practical framework for investigating how different choices of conditioning pixel subsets influence reconstruction outcomes.}

\subsection{Our Contribution and Paper Overview}

The main goal of this paper is to investigate the quality of full image reconstruction given only a subset of observed pixels\NK{---a setting that arises in several tasks, including image compression when only a small number of pixels are selected for storage \cite{superpixel_2023}}. Our tool for reconstruction is a generalized convolutional NADE model, \textit{ConvNADE-Beta-Color}, which we adapt for the first time to handle real-valued and color images (Section~\ref{sec:CONVBETAColor}). We also tailor the training procedure specifically for the image reconstruction task (Section~\ref{sec:new_way}). \NK{In particular, although we build upon the ConvNADE architecture, we depart from the traditional autoregressive formulation by conditioning on a structured subset of observed pixels. Our method is best described as a \textit{masked conditional density estimation} model built on NADE principles as described above. In Section~\ref{sec:numericalresults}, we evaluate this approach on several benchmark datasets, including Binarized MNIST, CIFAR-10, FER2013, and LHQ. We investigate whether our NADE-inspired model, when conditioned on a low-discrepancy pixel patch (see Section \ref{sec:new_way}), can achieve realistic reconstructions while remaining more efficient to train than modern high-capacity alternatives.}



\section{Background on NADE models}\label{sec:backgroundonNADE}

We begin by presenting the mathematical framework of the NADE models previously studied in the literature, following the notation established in \cite{larochelle2011neural,uria2016neural}. For comprehensive details, we refer the reader to these references.

\subsection{Neural Autoregressive Distribution Estimation (NADE)}

The Neural Autoregressive Distribution Estimation (NADE) model estimates high-dimensional probability distributions by factorizing the joint distribution \( p(\mathbf{x}) \) into a product of conditionals as in equation \eqref{eq:conditionals}. Each conditional probability \( p(x_{o_d} | \mathbf{x}_{o_{<d}}) \) is parameterized using a neural network with tied weights, ensuring the same parameters are shared across all conditionals. Let's assume for now that the dimensions of $\mathbf{x}$ are binary, i.e., $x_d \in \{0,1\}$ for $d \in \{1, 2, \ldots, D\}$. The hidden activations for the model are computed as
\[
\mathbf{h}_d = \sigma(\mathbf{W}_{\cdot,o_{<d}} \mathbf{x}_{o_{<d}} + \mathbf{c}),
\]
where \(\sigma(\cdot) \) is the logistic sigmoid function, \(\mathbf{W} \in \mathbb{R}^{H \times D}\) is the weight matrix (with $H$ as the number of hidden units), and \(\mathbf{c} \in \mathbb{R}^H\) is the bias vector. The conditional probabilities are given by
\[
p(x_{o_d} = 1 | \mathbf{x}_{o_{<d}}) = \sigma(\mathbf{V}_{o_d,\cdot} \mathbf{h}_d + b_{o_d}),
\]
where \(\mathbf{V} \in \mathbb{R}^{D \times H}\) and \(\mathbf{b} \in \mathbb{R}^D\) are trainable parameters. NADE is trained by minimizing the average negative log-likelihood of the data
\[
-\frac{1}{K} \sum_{k=1}^K \log p(\mathbf{x}^{(k)}) = -\frac{1}{K}\sum_{k=1}^K\sum_{d=1}^D \log p(x_{o_d}^{(k)}|\mathbf{x}_{o_{<d}}^{(k)})
\]
where 
$\mathbf{x}^{(k)}$ denotes the $k^{th}$ data point in the training set and $K$ is the total number of data points. Training employs stochastic gradient descent (or one of its variants) where gradients are efficiently computed in \(\mathcal{O}(HD)\) using a recurrence relation for the hidden layer pre-activations.

\subsection{DeepNADE}

DeepNADE extends NADE by introducing $L$ layers of hidden units in an effort to model more complex dependencies. The conditional probabilities are expressed as
\[
p(x_{o_d} | \mathbf{x}_{o_{<d}}) = \sigma(\mathbf{V}_{o_d}^\top \mathbf{h}_d^{(L)} + b_{o_d}),
\]
where \(\mathbf{h}_d^{(L)}\) is the activation of the topmost hidden layer, recursively computed as
\[
\mathbf{h}_d^{(l)} = \sigma(\mathbf{W}^{(l)} \mathbf{h}_d^{(l-1)} + \mathbf{c}^{(l)}), \quad l = 1, \dots, L.
\]

DeepNADE incorporates a binary mask \(\mathbf{M}_{o_{<d}}\) to ensure that predictions depend only on observed variables. The masked input is defined as
\[
\mathbf{h}_d^{(0)} = \mathbf{x} \odot \mathbf{M}_{o_{<d}},
\]
where \(\odot\) represents element-wise multiplication. The model is trained using an order-agnostic procedure that optimizes the expected log-likelihood over all orderings
\[
\mathcal{L}(\boldsymbol{\theta}) = -\frac{1}{K} \sum_{k=1}^K \mathbb{E}_{o \in \mathcal{D}!} \left[ \log p(\mathbf{x}^{(k)} |\boldsymbol{\theta}, o)  \right],
\]
where the models dependence on the network parameters $\boldsymbol{\theta}$ and the ordering $o$ are made explicit and $ \mathcal{D}!$ is the set of all ordered permutations of $\{1,2,\ldots,D\}$ (i.e. set of all orderings).
Importantly, the above formula for $\mathcal{L}(\boldsymbol{\theta})$ cannot be computed due to the very high number of terms required to determine the expected log-likelihood. In practice, the solution is to choose a subset of 128 random orderings from $\mathcal{D}!$ to approximate this expectation \cite{uria2016neural}, which still presents computational difficulties.

\subsection{Convolutional NADE (ConvNADE)}

ConvNADE adapts NADE specifically for image data by using convolutional layers, instead of fully connected layers, preserving the spatial structure of the data. Instead of a vector of observed data points, we now take $\mathbf{X} \in \{0,1\}^{N_\mathbf{X}\times N_\mathbf{X}}$ to denote a square binary image. The conditional probabilities are given by
\[
p(x_{o_d} =1 | \mathbf{x}_{o_{<d}}) = \textbf{vec}(\mathbf{H}^{(L)})_{o_d}
\]
where the notation $\textbf{vec}(\mathbf{X})$ refers to concatenation of every row in $\mathbf{X}$, and $\mathbf{H}^{(L)}$ is the output feature of the final convolutional layer. Similarly to DeepNADE, ConvNADE uses a binary mask \(\mathbf{M}_{o_{<d}}\), applied element-wise to the input
\[
\mathbf{H}^{(0)} = \mathbf{X} \odot \mathbf{M}_{o_{<d}}
\] 
and for any activation function $\sigma$, the feature maps are computed as
\[
\mathbf{H}_j^{(\ell)} = \sigma \left( \sum_{i=1}^{H^{(\ell-1)}} \mathbf{H}_i^{(\ell-1)} \circledast \mathbf{W}_{ij}^{(\ell)} + b_j^{(\ell)} \right), \quad \ell = 1,\ldots,L
\]
where $H^{(\ell)}$ is the number of feature maps output by layer $\ell$, $\mathbf{b}^{\ell} \in \mathbb{R}^{H^{(\ell)}}$ is the bias for each layer $\ell$ and \(\mathbf{W}_{ij}^{(\ell)} \in \mathbb{R}^{N^{\ell}_W \times N^{\ell}_W}\) are square convolutional filters of size $N_W^{(\ell)}$ varying for each layer $\ell$ connecting two feature maps $\mathbf{H}_i^{(\ell-1)}$ and $\mathbf{H}_j^{(\ell)}$.  Following \cite{uria2016neural}, the notation $\circledast$ is used to denote both ``valid'' convolutions and ``full'' convolutions where ``valid'' convolutions only apply a filter to complete patches of the image, resulting in a smaller image. For ``full'' convolutions, the image contour is zero-padded before applying the convolution, resulting in a larger image. Note that these convolutions use a stride of 1.

ConvNADE is trained using the same order-agnostic procedure as DeepNADE
and therefore possesses the same intractability issue when computing the expectation.

\section{A Generalized ConvNADE Model}\label{sec:CTS}

A limitation of the original NADE and ConvNADE models is their reliance on Bernoulli distributions for conditional probabilities, restricting their applicability to binary datasets. For problems involving real-valued response distributions, the RNADE model \cite{uria2013rnade} is commonly employed, utilizing a mixture of Gaussian distributions to model the response variable. However, this approach requires training a substantial number of parameters---twice the number of mixture components---which could make the model expensive to fit. As highlighted in \cite{uria2013rnade}, while Gaussian mixtures are a popular choice for modeling continuous responses, alternative distributions may better suit specific tasks.


In this text, we extend the ConvNADE model discussed in Section \ref{sec:backgroundonNADE} to handle color datasets, broadening its utility. Moreover, we demonstrate that using a single Beta distribution to model the response variable is not only feasible but also capable of producing clear, realistic outputs. This approach is more computationally efficient requiring the training of only two parameters, the shape and scale of the Beta distribution, \NK{however, despite not seen in practice with any of the Section \ref{sec:numericalresults} experiments, we acknowledge that the uni-modal nature of the Beta distribution could limit its expressiveness when the true conditional distribution is multi-modal.}

\subsection{ConvNADE-Beta-Color}\label{sec:CONVBETAColor}
We adapt the ConvNADE model by modeling the conditional distribution of each color channel of an input image as a Beta distribution while also adding in color channels. We call this variant \textit{ConvNADE-Beta-Color}.

The ConvNADE-Beta-Color model outputs six channels: red, green and blue color channels, each parameterized by two Beta distribution parameters (shape and scale). Importantly, the model structure remains consistent from layer 1 to layer $L-1$ with modifications applied only to the final output layer. For a color input image $\mathbf{X}$ of dimension $(3,N_\mathbf{X},N_\mathbf{X})$ the initial input $\mathbf{H}^{(0)}$ to the model is produced by elementwise-multiplication with the concatenated binary mask
\[
\mathbf{H}^{(0)} = \mathbf{X} \odot \text{concat}(\mathbf{M}_{o_{<d}}, \mathbf{M}_{o_{<d}}, \mathbf{M}_{o_{<d}}).
\]

\begin{figure}[t]
    \centering
    \includegraphics[width=0.4\linewidth]{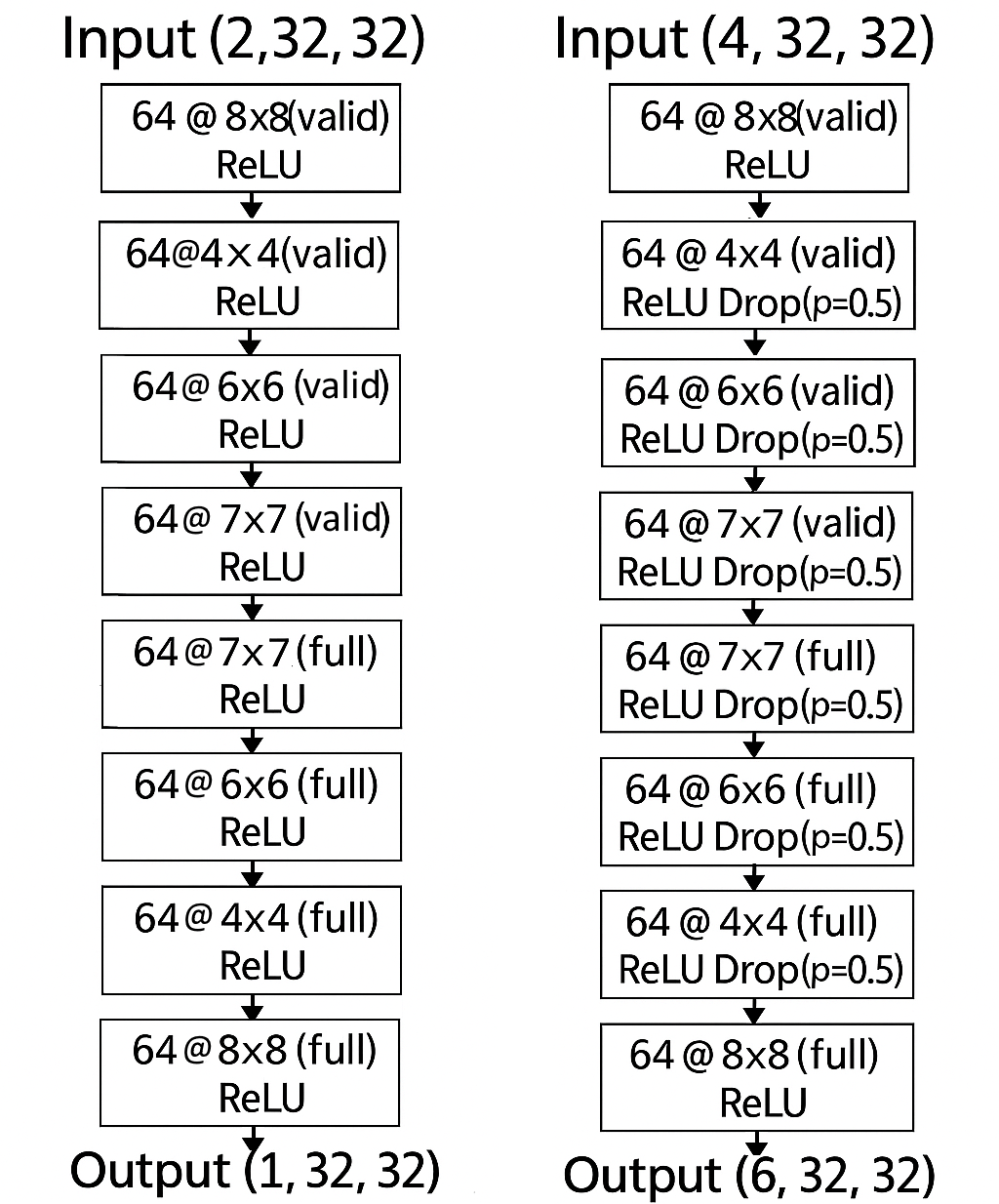}
    \caption{ConvNADE and ConvNADE-Beta-Color with $L=8$ convolutional layers, each defined as \texttt{channels @ filter size (conv type)}. For ConvNADE, the image has 2 channels: one for the binary pixel value and one for the mask, and outputs a 1 layer Bernoulli parameter map. For ConvNADE-Beta-Color, the image has 4 channels: one for each color and one for the mask and the output channel has $6$ layers representing Beta shape and scale for each color.}
    \label{fig:convnade}
\end{figure}

The output layer $L$ computes two parameters for each color channel, corresponding to the shape and scale parameters of the Beta distribution. This is computed as follows
\[
\mathbf{A}^{(L)}_{2c-1} = \sum_{i=1}^{H^{(L-1)}} \mathbf{H}^{(L-1)}_i \circledast \mathbf{W}^{(L)}_{i(2c-1)} +b^{(L)}_{2c-1} , \quad
\mathbf{H}^{(L)}_{2c-1} = \log(1 + e^{\mathbf{A}^{(L)}_{2c-1}})
\]

\[
\mathbf{A}^{(L)}_{2c} = \sum_{i=1}^{H^{(L-1)}} \mathbf{H}^{(L-1)}_i \circledast \mathbf{W}^{(L)}_{i(2c)} + b^{(L)}_{2c} , \quad
\mathbf{H}^{(L)}_{2c} = \log(1 + e^{\mathbf{A}^{(L)}_{2c}})
\]
where \(c = 1, 2, 3\) corresponds to the red, green, and blue color channels and $H^{(L-1)}$ denotes the number of channels in layer $L-1$. The weight matrix in the $L^{th}$ layer is denoted $\mathbf{W}^{(L)}\in\mathbb{R}^{H^{(L-1)}\times 6\times  N_\mathbf{X}\times N_\mathbf{X}}$ and $b^{(L)}\in\mathbb{R}^{6}$ is the bias vector in the $L^{th}$ layer. The final conditional distribution for each channel is then modeled as
\[
p(\textbf{vec}(\mathbf{X}_c)_{o_d} \mid \textbf{vec}(\mathbf{X}_c)_{o_{<d}}) = p_{\text{Beta}}(\textbf{vec}(\mathbf{X}_c)_{o_d}; \textbf{vec}(\mathbf{H}^{(L)}_{2c-1})_{o_d}, \textbf{vec}(\mathbf{H}^{(L)}_{2c})_{o_d}),
\]
for each $c=1,2,3$. Figure \ref{fig:convnade} presents a simple schematic of the ConvNADE-Beta-Color model where we ensure the first half of its layers uses ``valid'' convolutions while the other half uses ``full'' convolutions. It was found that including the mask $\mathbf{M}_{o_{< d}}$ as an input to the model improves performance \cite{uria2016neural}, as it resolves ambiguity about whether a value is $0$ or has been masked to $0$.
For ConvNADE-Beta-Color, the mask was provided as an additional channel to the input layer hence the four channel input in Figure \ref{fig:convnade}.

\subsection{Training ConvNADE-Beta-Color for Image Reconstruction}\label{sec:new_way}

\NK{As previously noted, ConvNADE models are traditionally trained on a random subset of pixel orderings sampled from the full set of permutations, $\mathcal{D}!$. This corresponds to conditioning on one pixel at a time in a fully autoregressive manner. However, in many practical applications, such as image inpainting and data compression, the `order' of pixels is not explicitly known and it may be realistic to condition on a fixed subset of observed pixels \cite{inpainting2010,inpainting2022,image_compress_2006}. In this section, we depart from the standard NADE objective of modeling the full joint distribution via a chain of one-dimensional conditionals. Instead, we propose a method to train the ConvNADE-Beta-Color model that learns to predict missing pixels given a pre-defined subset. While the architecture, including its use of masking, remains recognizable as derived from its NADE parent, the training procedure is adapted for fixed-pattern conditioning rather than full autoregression.}

\begin{figure}[t]
    \centering
    \includegraphics[width=0.5\textwidth]{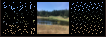}
    \label{fig:patches}

\caption{A $32 \times 32$ image from LHQ illustrating pixel patches. (\textbf{Left}) A random patch and (\textbf{Right}) a low-discrepancy patch of size 128.}
\label{fig:patches}
\end{figure}

\subsubsection{Pixel Patches.}
Given an image $\mathbf{X}$ containing $D$ pixels, index each pixel from $\{1,\ldots,D\}$ from the top left most pixel to bottom right, row-wise. We address the problem of reconstructing the image when observing only a small subset of pixels of size $P < D$ referred to as a \textit{pixel patch of size $P$}; see Figure \ref{fig:patches}. That is, our goal is to accurately predict the remaining $D-P$ pixels of the image. 

In the following, we describe two pixel patches that will be investigated. A \textit{random pixel patch of size $P$} is defined as a subset \( \mathcal{S}^{\text{rand}}_P \subseteq \{1,\ldots,D\}\) sampled uniformly at random from the set of all subsets of \( \{1, \ldots,D\} \), i.e.,
$\mathcal{S}^{\text{rand}}_P \sim U\big(\{S \subseteq \{1, \ldots, D\} : |S| = P\}\big).$
Geometrically, this corresponds to selecting \( P \) pixels uniformly at random without replacement from the entire pixel grid.

In contrast, a \textit{low-discrepancy pixel patch} can be constructed deterministically only for power of integer $b$ number of points. For standardization purposes across datasets, all images used in our experiments in Section~\ref{sec:numericalresults} are rescaled to \( 2^{5} \times 2^{5} \). With this in mind, we define a low-discrepancy pixel patch for images of dimension \( 2^m \times 2^m \) for \( m \in \mathbb{N} \). Thus, let \( \{u_n = (u_n^{(1)}, u_n^{(2)}) \}_{n=1}^{2^{2m}} \) denote the first \( 2^{2m} \) points of the two-dimensional Sobol' sequence \cite{SOBOL196786}. A low-discrepancy pixel patch of size $2^k$ for $k \leq m$ is defined as a subset $\mathcal{S}^{\text{LD}}_P \subseteq \{1,\ldots,D\}$ containing the indices
\[
\left\{\left\lfloor 2^m \cdot u_n^{(1)} \right\rfloor + 2^m \left( 2^m - 1 - \left\lfloor 2^m \cdot u_n^{(2)} \right\rfloor \right) +1 \right\}_{n=1}^{2^{k}}.
\]
This mapping leverages the stratification of the Sobol' sequence, an example of a popular quasi-Monte Carlo construction called a digital net \cite{dick2010digital}, to create a uniformly distributed subset of pixels.

\subsubsection*{Objective Function.}
We now design the loss function to be implemented in ConvNADE-Beta-Color for the purpose of image reconstruction. Let $\mathcal{X}$ be the training dataset containing $K$ real-valued images $\mathbf{X} \in [0,1]^{N_\mathbf{X} \times N_{\mathbf{X}}}$ of dimension $(3, N_\mathbf{X}, N_\mathbf{X})$, where the first three channels are for the colors and an additional channel is to be added for the mask. The dataset is organized into $B$ batches with a batch size of $m$ (i.e. $K=B\times m$). Note that $D = N_\mathbf{X} \times N_\mathbf{X}$. We denote the $c^{th}$ color channel of $i^{th}$ image in the $b^{th}$ batch by $\mathbf{X}^{(b)}_{i,c}$ for $c = 1, 2, 3$, $i = 1, 2, \ldots, m$ and $b = 1, 2, \ldots, B$. 
Assuming $\boldsymbol{\alpha}_c =  \textbf{vec}(\mathbf{H}^{(L)}_{2c-1})$ and $\boldsymbol{\beta}_c =  \textbf{vec}(\mathbf{H}^{(L)}_{2c})$ are the flattened outputs of the two channels for each color channel $c$, to train ConvNADE-Beta-Color we optimize the following loss function for each mini-batch $b = 1,2,\ldots,B$
\begin{equation*}\label{eqn:patch_loss}
    \mathcal{L}_{b,P, \text{Beta}}(\boldsymbol{\theta}) = -\frac{1}{3m(D-P)}\sum_{i=1}^{m}\sum_{c=1}^{3}\sum_{j\not\in \mathcal{S}^{\bullet}_{P}}\log p_{\text{Beta}}(\textbf{vec}(\mathbf{X}^{(b)}_{i,c})_{j}| \boldsymbol{\alpha}, \boldsymbol{\beta},\textbf{vec}(\mathbf{X}^{(b)}_{i,c})_{\mathcal{S}^{\bullet}_{P}}),
\end{equation*}
where $\bullet \in \{\text{rand}, \text{LD}\}$ depending on type of pixel patch implemented. 
To circumvent over-fitting, between each hidden layer, we perform a dropout (see \cite{srivastava2014dropout}) with probability 50\%.

\section{Results}\label{sec:numericalresults}

The primary objectives of this study are twofold: first, to evaluate the ability of the proposed ConvNADE-Beta-Color model to perform image reconstruction, and second, to assess how conditioning on different observed pixel patches impacts the model's performance. That is, can ConvNADE-Beta-Color reconstruct clear and coherent images given a partially observed image? We answer this question in the affirmative as demonstrated in Appendix \ref{app:visuals}. Additionally, we examine the effect of conditioning on random versus low-discrepancy pixel patches, following the methodology outlined in Section \ref{sec:new_way}.  

To evaluate performance, we train the ConvNADE-Beta-Color model on benchmark datasets, including FER2013 \cite{goodfellow2013challenges}, LHQ \cite{skorokhodov2021aligning}, CIFAR-10 \cite{krizhevsky2009learning}, and Binarized MNIST \cite{salakhutdinov2008quantitative}. For Binarized MNIST, we use the original ConvNADE architecture described in Section \ref{sec:backgroundonNADE}, incorporating a similarly adapted training procedure as detailed in Section \ref{sec:new_way} for ConvNADE-Beta-Color relating to the pixel patches. During testing, the model is conditioned on pixel patches consistent with the training setup and tasked with predicting the missing pixels. Reconstruction performance is evaluated using test loss values and qualitative image quality, comparing results across different patch types.


\subsection{Training Details} 
All experiments have been run on NVIDIA GeForce RTX 4050. Models were trained using a pixel patch size of 128 and for either 60 or 120 epochs depending on the point in which the training loss began to plateau. The batch size, $m$, was fixed at 100 for all experiments. We used the Adam optimizer \cite{kingma2014adam} with a learning rate of $10^{-4}$ which was found to be the optimal choice for training.
The deterministic Sobol' sequence required for the low-discrepancy pixel patch was generated via the \texttt{qmcpy} python package \cite{qmcpy_doc}. All images are resized to $32\times32$. 

\subsection{Datasets}

We evaluate our models on several classical datasets; Binarized MNIST, FER2013, CIFAR-10, and LHQ. Binarized MNIST serves as a classical benchmark containing images of handwritten digits from $0$ to $9$. We use the standard dataset split, consisting of 50000 training examples, 10000 validation examples, and 10000 test examples. The FER2013 dataset contains centered greyscale facial images with varying expressions. Originally sized at \(48 \times 48\) pixels, these images are resized to \(32 \times 32\) pixels to facilitate training. The dataset is divided into 24403 training examples, 4306 validation examples, and 3589 test examples. The CIFAR-10 dataset includes \(32 \times 32\) color images of 10 object classes, such as trucks, dogs, and ships, with 6000 images per class. For our experiments, we use a train-validation-test split of 44000 training examples, 6000 validation examples, and 10000 test examples. Finally, the LHQ dataset consists of colored landscape images originally sized at \(256 \times 256\) pixels, which are scaled down to \(32 \times 32\) pixels for training purposes. The dataset is divided into 63000 training examples, 13500 validation examples, and 13500 test examples.

\begin{figure}[t]
\captionsetup[subfigure]{labelformat=empty}
\centering
\begin{subfigure}{0.49\textwidth}
    \centering
    \includegraphics[width=\textwidth]{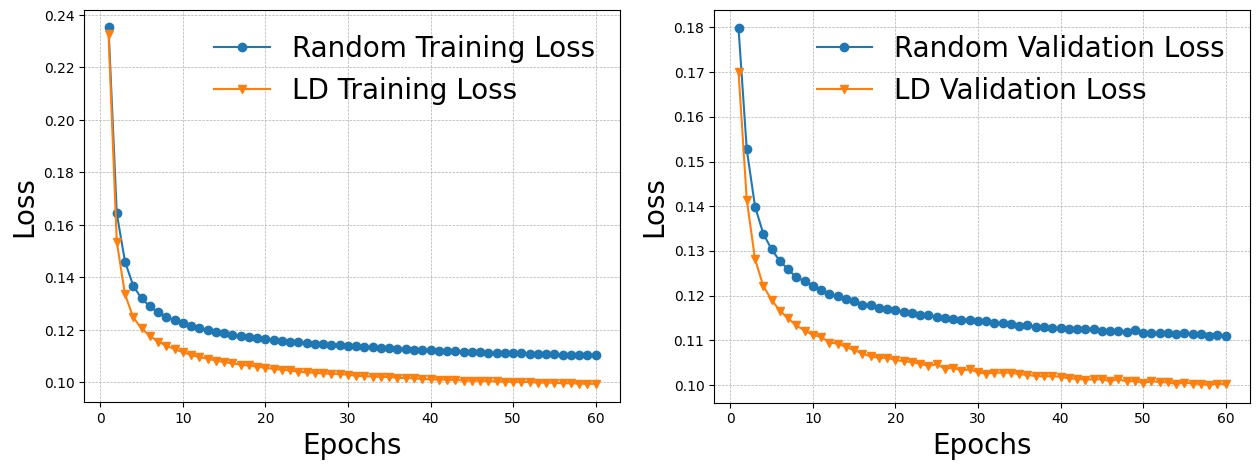}
    \caption{Binarized MNIST}
    \label{fig:binmnist_curve}
\end{subfigure}
\hfill
\begin{subfigure}{0.49\textwidth}
    \centering
    \includegraphics[width=\textwidth]{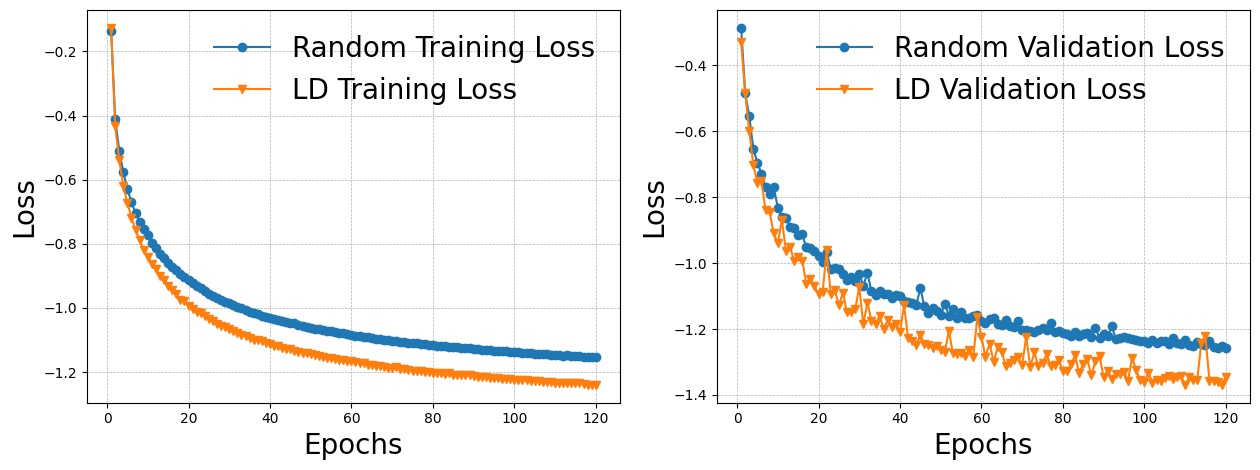}
    \caption{FER2013}
    \label{fig:fer2013_curve}
\end{subfigure}

\vspace{0.5cm}

\begin{subfigure}{0.49\textwidth}
    \centering
    \includegraphics[width=\textwidth]{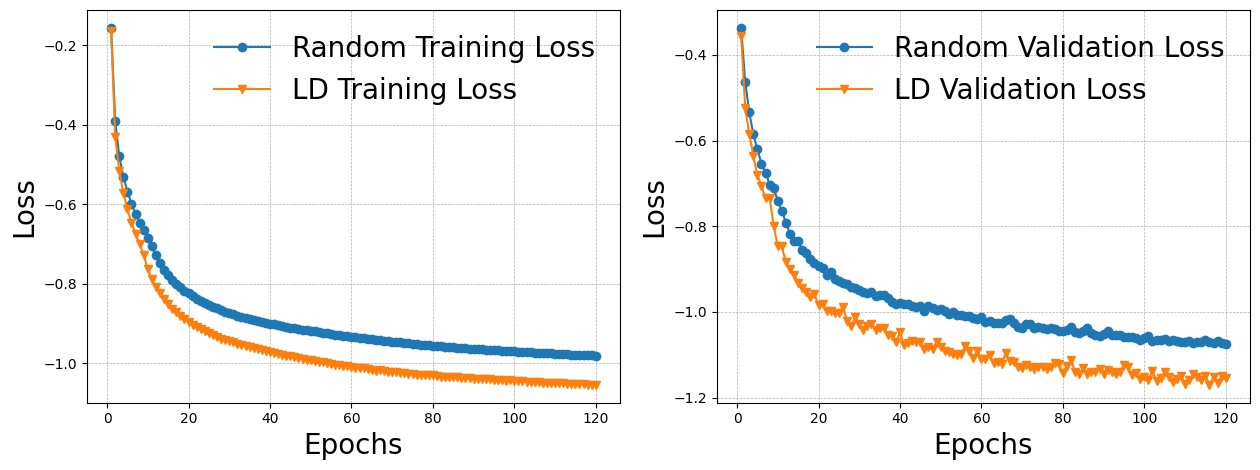}
    \caption{CIFAR-10}
    \label{fig:cifar_curve}
\end{subfigure}
\hfill
\begin{subfigure}{0.49\textwidth}
    \centering
    \includegraphics[width=\textwidth]{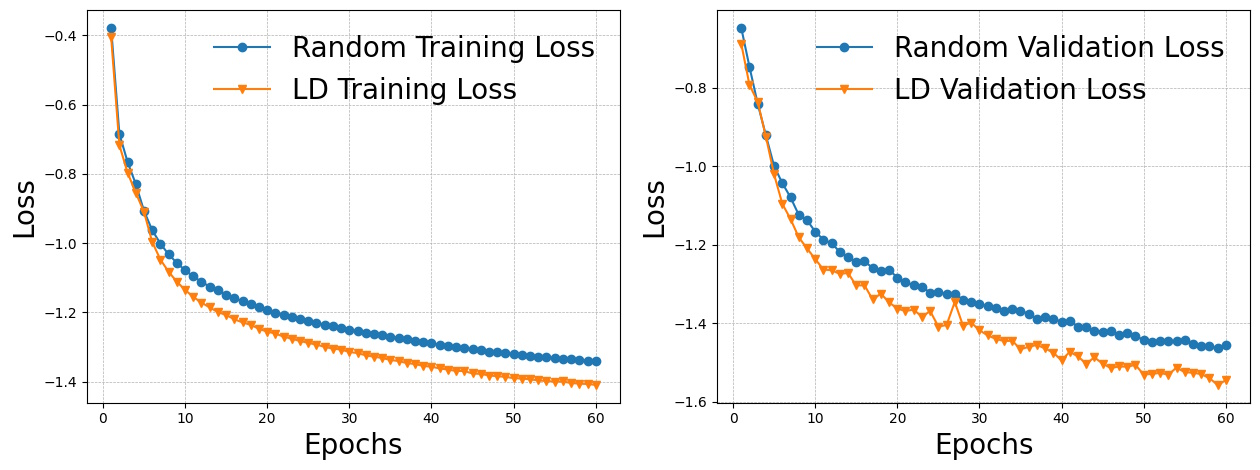}
    \caption{LHQ}
    \label{fig:LHQ_curve}
\end{subfigure}
\caption{For each data set, the left and right panels show the training and validation loss for random and low-discrepancy pixel patches. For the random patch, the mean loss over five random orderings is given.}
\label{fig:loss_curves}
\end{figure}

\subsection{Numerical Results}

Table \ref{tab:test_loss} presents the test loss values for both the ConvNADE and ConvNADE-Beta-Color models when trained on random and low-discrepancy pixel patches across several datasets. For the random patches, the value returned is the mean average loss over five random patches with $95\%$ confidence intervals provided.

Consistently across all experiments, we observe that training with low-discrepancy patches results in significantly lower test loss values compared to random patches. Figure \ref{fig:loss_curves} further illustrates that the training and validation curves for low-discrepancy patches converge more quickly and to lower loss values than those for random patches. Qualitative results, shown in Appendix \ref{app:visuals}, demonstrate that reconstructions using low-discrepancy patches are sharper and display more detailed object features. 

\AEW{We also compared training times of ConvNADE-Beta-Color with ConvNADE-MoG-Color on the FER2013 dataset. ConvNADE-MoG-Color is otherwise identical to ConvNADE-Beta-Color, as described in Section~\ref{sec:CONVBETAColor}, except that it uses a mixture of Gaussians (in the case of this experiment, a mixture of $10$ Gaussians) to model the output distribution for each color channel. We can report that ConvNADE-Beta-Color completed training in 2375 seconds, faster than ConvNADE-MoG-Color, which required 3221 seconds across 120 epochs. In addition to this improved efficiency, ConvNADE-Beta-Color avoids the need to select the number of mixture components, a hyperparameter required by ConvNADE-MoG-Color.}

\begin{table}[t]
\centering

\begin{tabular}{lll}
\hline
Patches & \vline \hspace{1mm} Bin. MNIST & FER2013 \\ \hline
Random & \vline\hspace{1mm} $0.11$ $(0.102, 0.119)$ & $-1.252$ $(-1.257, -1.247)$ \\
LD     & \vline\hspace{1mm} $\mathbf{0.0994}$ & $\mathbf{-1.3367}$ \\ \hline
\end{tabular}

\vspace{0.2cm} 

\begin{tabular}{lll}
\hline
Patches & \vline \hspace{1mm} CIFAR-10 & LHQ \\ \hline
Random & \vline\hspace{1mm} $-1.077$ $(-1.086, -1.068)$ & $-1.456$ $(-1.484, -1.429)$ \\
LD     & \vline\hspace{1mm} $\mathbf{-1.1536}$ & $\mathbf{-1.5468}$ \\ \hline
\end{tabular}

\vspace{2mm}
\caption{Test loss results for the ConvNADE model on the Binarized MNIST dataset, and for the ConvNADE-Beta-Color model on FER2013, CIFAR-10, and LHQ datasets. Results for random patches are mean average loss over 5 random patches with $95\%$ CI.}
\label{tab:test_loss}
\end{table}

\section{Discussion}
In this paper, we propose a variant of the convolutional neural autoregressive distribution estimator, referred to as \textit{ConvNADE-Beta-Color}, designed specifically for reconstructing images from a limited subset of observed pixels. This model extends the original ConvNADE framework to accommodate continuous and color images, significantly broadening its applicability, while having the benefit of being much less computationally expensive than its original counterparts. Through experiments across multiple benchmark datasets, we analyze the impact of different input pixel patches. Our results consistently show that ConvNADE-Beta-Color produces coherent reconstructions and further, observing a low-discrepancy pixel patch leads to lower test loss, faster convergence, and clearer image reconstructions versus observing a randomly selected patch. 

\NK{Although in many practical settings the observed pixels are fixed by the acquisition process, our results suggest that whenever there is flexibility in which pixels are observed, choosing them with uniformity is beneficial. This principle is relevant in systems where pixel access can be controlled such as single-pixel cameras \cite{Duarte_2008} and superpixel-based image compression \cite{superpixel_2023} where the observation pattern can be explicitly designed.}

Future work could explore extending this approach to color image generation and applying low-discrepancy pixel patches to enhance image compression tasks.

\bibliographystyle{spmpsci}
\bibliography{refs.bib}

\paragraph{Notes and Comments.}
The first author is supported by NSERC Discovery Grant RGPIN-238959. The second author is supported by NSF-DMS Grant \#2316011. The authors would like to thank Christiane Lemieux for helpful conversations throughout the development of this work. Our model, datasets and python code to reproduce results are available at:

\hyperlink{https://www.github.com/AmbroseEmmettIwaniw/}{GitHub/AmbroseEmmettIwaniw}. 

\appendix

\section{Image reconstructions from Section 4 Experiments}\label{app:visuals}


\begin{figure}[h!]
     \centering
     \includegraphics[width=0.5\textwidth]{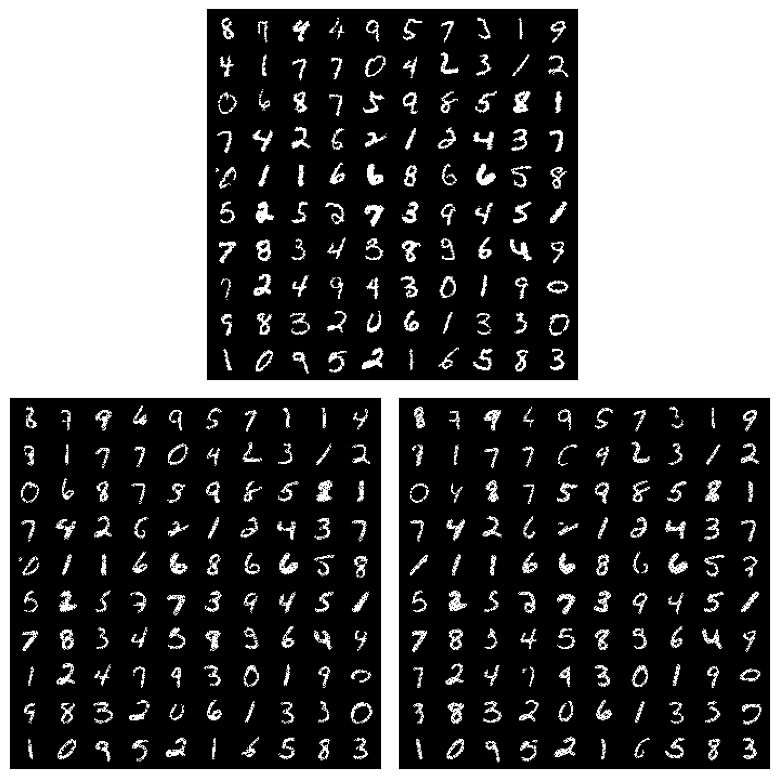}
    \caption{Reconstructed images from binarized MNIST using the ConvNADE model. (\textbf{Top}) Original binarized MNIST images randomly sampled. (\textbf{Bottom-Left}) Reconstructed with random patch. (\textbf{Bottom-Right}) Reconstructed with low-discrepancy patch.}
    \label{fig:mnist_recon}
\end{figure}

\begin{figure}[h!]
     \centering
     \includegraphics[width=0.5\textwidth]{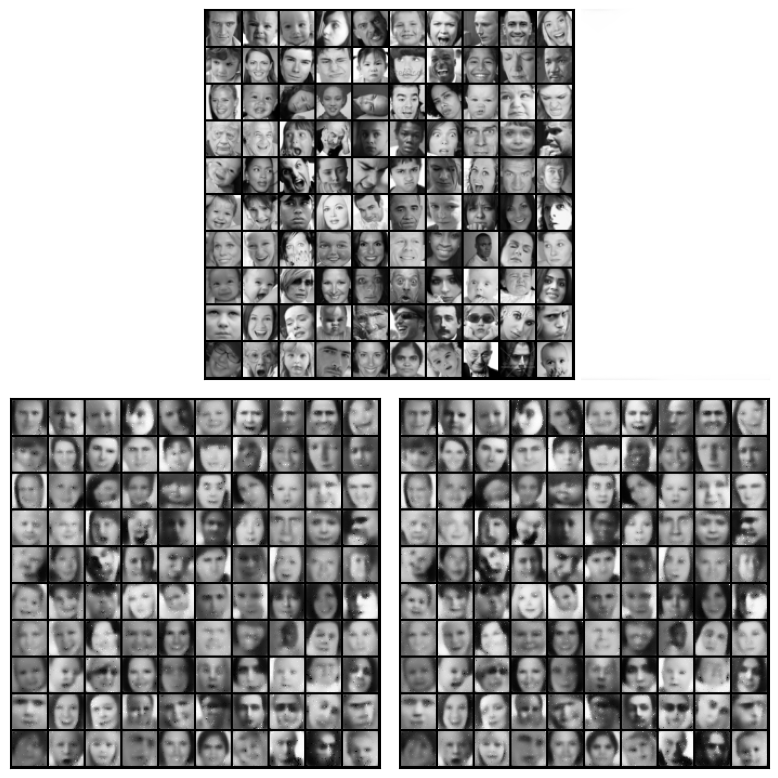}
    \caption{Reconstructed images from FER2013 using the ConvNADE-Beta-Color model (\textbf{Top}) Original FER2013 images randomly sampled. (\textbf{Bottom-Left}) Reconstructed with random patch. (\textbf{Bottom-Right}) Reconstructed with low-discrepancy patch.}
    \label{fig:fer_recon}
\end{figure}

\begin{figure}
  \centering
    \includegraphics[width=0.55\textwidth]{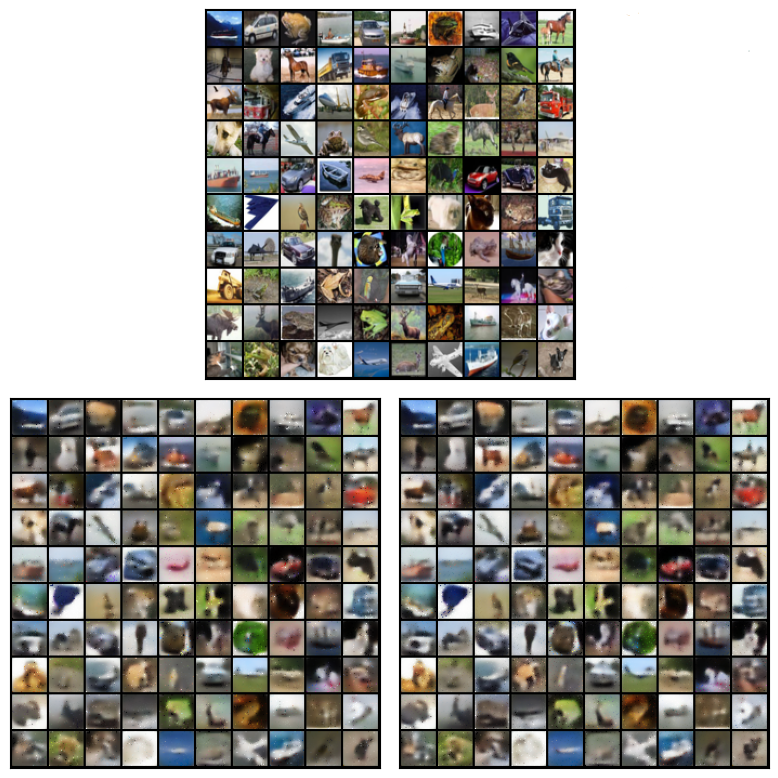}
    
    \caption{Reconstructed images from CIFAR-10 using the ConvNADE-Beta-Color model (\textbf{Top}) Original CIFAR-10 images randomly sampled. (\textbf{Bottom-Left}) Reconstructed with random patch. (\textbf{Bottom-Right}) Reconstructed with low-discrepancy patch.}
    \label{fig:cifar_recon}
\end{figure}

\begin{figure}
     \centering
     \includegraphics[width=0.55\textwidth]{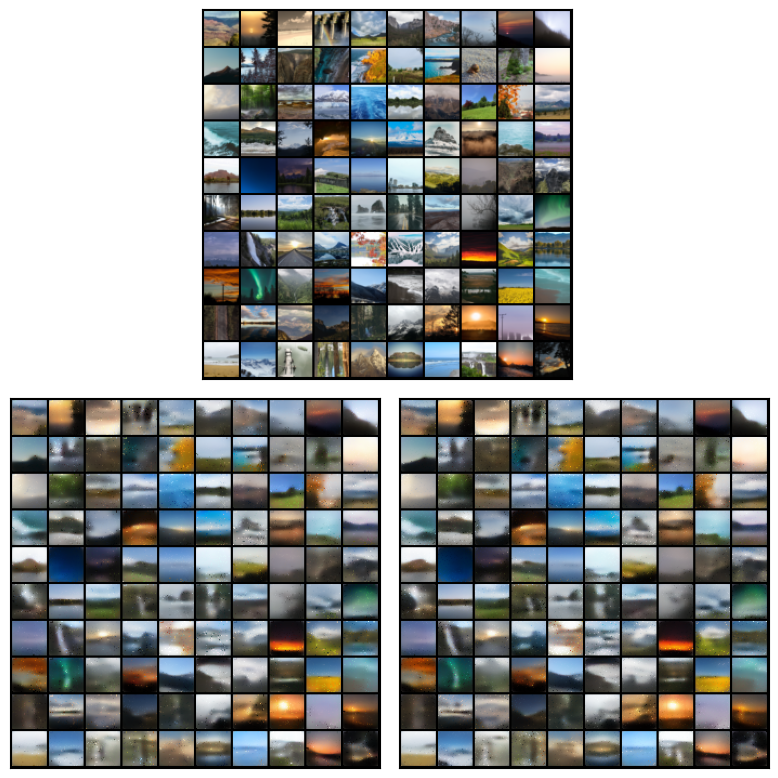}
    
    \caption{Reconstructed images of LHQ using the ConvNADE-Beta-Color model (\textbf{Top}) Original LHQ images randomly sampled. (\textbf{Bottom-Left}) Reconstructed with random patch. (\textbf{Bottom-Right}) Reconstructed with low-discrepancy patch.}
    \label{fig:lhq_recon}
\end{figure}

%
%

%

\end{document}